\documentclass[aps,a4paper,twocolumn]{revtex4-2}
\usepackage{amsmath}
\usepackage{graphicx}
\usepackage{subfigure}
\usepackage[usenames,dvipsnames]{color}
\definecolor{darkblue}{RGB}{0,0,196}
\usepackage[colorlinks=true,linkcolor=darkblue,citecolor=darkblue,urlcolor=darkblue]{hyperref}
\usepackage{setspace}
\usepackage{hyperref}
\usepackage{xcolor}
\usepackage{booktabs}
\hypersetup{
  colorlinks   = true, 
  urlcolor     = red, 
  linkcolor    = blue, 
  citecolor   = blue 
}
\usepackage{graphicx}
\usepackage{amsmath,bbm}
\usepackage{amssymb,bm}
\usepackage{yfonts}
\usepackage{comment}

\usepackage[usenames,dvipsnames]{color}
\definecolor{darkblue}{RGB}{0,0,196}
\usepackage[colorlinks=true,linkcolor=darkblue,citecolor=darkblue,urlcolor=darkblue]{hyperref}

\begin{document}
\title{Pinning down the origin of ``strangeness enhancement'' in the jet-like region of pp collisions with Monte Carlo simulations}

\author{Antonio Ortiz}
\author{Dushmanta Sahu}
\email{Dushmanta.Sahu@cern.ch}
\author{Victor Vázquez}
\affiliation{Instituto de Ciencias Nucleares, UNAM, Apartado Postal 70-543, Coyoacán, 04510, México City, México}

\begin{abstract}
A recent ALICE measurement reports a finite strangeness enhancement in the toward-leading region of high-multiplicity proton-proton collisions. In this work, the trigger transverse momentum, $p_{\rm T}^{\rm trigg}$, dependence is investigated using ${\rm K}^{0}_{\rm S}$ and $\Xi^{\pm}$ production in PYTHIA8, with and without color ropes, and EPOS-LHC. For $3 < p_{\rm T}^{\rm trigg} < 15$~GeV/$c$, EPOS-LHC and PYTHIA8 with color ropes exhibit a multiplicity-dependent increase of the toward-leading $\Xi^{\pm}/{\rm K}^{0}_{\rm S}$ ratio, while the rope-free PYTHIA8 baseline shows a weaker dependence. With increasing trigger momentum, the enhancement is strongly suppressed in EPOS-LHC for $p_{\rm T}^{\rm trigg}\gtrsim10$~GeV/$c$, whereas it remains approximately unchanged in PYTHIA8 with color ropes. This contrasting behavior is consistent with a reduced core contribution in EPOS-LHC and a local string-overlap mechanism in PYTHIA8 with color ropes. The trigger transverse momentum thus provides a model-sensitive probe of the mechanisms contributing to the apparent strangeness enhancement in the jet-like region of the pp collisions.

\end{abstract}

\maketitle

\section{Introduction}

The enhanced production of strange hadrons relative to non-strange hadrons is one of the most prominent signatures associated with the formation of a deconfined quark--gluon plasma (QGP)~\cite{Rafelski:1982pu,Koch:1986ud}. Originally proposed as a consequence of abundant strange-quark production in a thermalized partonic medium, strangeness enhancement was first observed in nucleus--nucleus collisions at the SPS~\cite{WA97:1998bwp,NA57:2006aux} and later confirmed at RHIC and the LHC~\cite{STAR:2007cqw,ALICE:2013xmt}. More recently, measurements performed by the ALICE Collaboration revealed that the yields of strange and multi-strange hadrons increase continuously with charged-particle multiplicity not only in heavy-ion collisions but also in high-multiplicity pp and p--Pb collisions~\cite{ALICE:2016fzo}. The observation of a common multiplicity scaling across collision systems of vastly different sizes has raised fundamental questions about the microscopic origin of strangeness production. The emergence of other collective-like phenomena in small systems, including long-range azimuthal correlations and finite flow coefficients, has further strengthened the analogy between high-multiplicity pp collisions and heavy-ion collisions~\cite{CMS:2010ifv,ALICE:2013snk,CMS:2016fnw}. At the same time, several microscopic mechanisms unrelated to QGP formation, such as color reconnection, color ropes, string interactions, and parton transport dynamics, have been shown to generate qualitatively similar signatures~\cite{OrtizVelasquez:2013ofg,Bierlich:2014xba,Lin:2004en}. Disentangling genuine collective effects from alternative mechanisms associated with event activity therefore remains a central challenge in the study of small-collision systems.
A particularly powerful approach is to investigate strange-hadron production relative to a high-$p_{\mathrm T}$ trigger particle. Such measurements provide a natural separation between particle production associated with hard scatterings and jet fragmentation, and particle production arising from the underlying event (UE)~\cite{ALICE:2019mmy}. Previous ALICE measurements demonstrated that the multiplicity-dependent enhancement of strange hadrons is significantly larger in UE-dominated regions than in jet-like regions, indicating that the underlying event plays a major role in the observed enhancement~\cite{ALICE:2022ecr}.

Nevertheless, a recent angular-correlation study in ALICE revealed that a finite enhancement persists even in the toward-leading region, commonly associated with jet fragmentation~\cite{ALICE:2024zxp}. This raises an important question: does the enhancement in the jet-like region reflect a modification of jet fragmentation, or does it arise from residual medium-like contributions associated with the underlying event? A critical aspect is the trigger-particle selection. The ALICE analysis employed leading charged particles with $p_{\mathrm T}^{\rm trigg}>3$~GeV/$c$, a threshold chosen to maximize statistical precision. However, particles in this momentum range can still receive substantial contributions from the underlying event (multiparton interactions), so the toward-leading region may not correspond to a purely fragmentation-dominated environment. Recently, one of our studies has shown that medium-like effects remain significant up to trigger momenta of approximately 6--8~GeV/$c$ in pp collisions, above which the selected particle population becomes increasingly dominated by hard processes~\cite{Sahu:2025mla}. This motivates a systematic investigation of the trigger-momentum dependence of the reported strangeness enhancement.

The novelty of the current study is to systematically vary the minimum transverse momentum of the trigger particle while keeping the correlation and multiplicity selections fixed. This allows us to test whether the multiplicity-dependent $\Xi^\pm/{\rm K}^{0}_{\rm S}$ enhancement observed in the toward-leading region persists as increasingly hard trigger particles are selected. Using PYTHIA8 with and without color-rope formation~\cite{Sjostrand:2014zea,Bierlich:2014xba} and EPOS-LHC with parametrized hydrodynamic evolution~\cite{Pierog:2013ria}, the strange-hadron production as a function of the trigger momentum threshold is investigated. By progressively increasing $p_{\mathrm T}^{\rm trigg}$ beyond the value employed in the ALICE analysis~\cite{ALICE:2024zxp}, increasingly hard trigger populations are selected, allowing a systematic investigation of the trigger dependence of the reported strangeness enhancement. Establishing whether the enhancement persists under increasingly hard trigger selections therefore provides a discriminating test of the mechanisms underlying strangeness enhancement in small-collision systems.

This paper is organized as follows. In section~\ref{sec2}, the methodology of the analysis is discussed along with the usage of event generators. Section~\ref{sec3} discusses the results of the analysis, and finally the conclusion and outlook are presented in section~\ref{sec4}.

\section{Methodology}
\label{sec2}

\subsection{Analysis framework}

The present analysis follows the strange-hadron correlation method introduced by the ALICE Collaboration~\cite{ALICE:2024zxp}. The trigger particle is defined as the highest-$p_{\rm T}$ charged particle in the event satisfying $|\eta|<0.8$, considering $\pi^\pm$, ${\rm K}^\pm$, ${\rm p}/\bar {\rm p}$, ${\rm e}^\pm$, and $\mu^\pm$. To investigate the dependence of strangeness enhancement on the hardness of the selected event, several trigger-momentum thresholds are considered, starting from the nominal ALICE selection of $p_{\rm T}^{\rm trigg}>3$~GeV/$c$ and extending to progressively higher values. Correlations between the trigger particle and associated ${\rm K}^0_{\rm S}$ and $\Xi^\pm$ hadrons are constructed in terms of the pseudorapidity and azimuthal-angle differences,
\begin{equation}
\Delta\eta = \eta_{\rm assoc}-\eta_{\rm trigg},
\qquad
\Delta\varphi = \varphi_{\rm assoc}-\varphi_{\rm trigg}.\nonumber
\end{equation}

Three correlation regions are considered. The toward-leading region is defined by $|\Delta\eta|<0.86$ and $|\Delta\varphi|<1.1$, and is predominantly associated with the trigger direction and contributions from hard processes and jet fragmentation. The transverse-to-leading region is defined by $0.86<|\Delta\eta|<1.2$ and $0.96<\Delta\varphi<1.8$, and is primarily used to characterize particle production in the underlying event. The full region is defined by $|\Delta\eta|<1.2$ and $-\pi/2<\Delta\varphi<3\pi/2$, encompassing both the toward-leading and transverse-to-leading regions.

The per-trigger yield of associated strange hadrons is defined as
\begin{equation}
Y(p_{\rm T})=
\frac{1}{N_{\rm trigg}}
\frac{1}{\Delta\eta\Delta\varphi}
\frac{{\rm d}N_{\rm assoc}}{{\rm d}p_{\rm T}},
\end{equation}
where $N_{\rm trigg}$ denotes the number of selected trigger particles and ${\rm d}N_{\rm assoc}/{\rm d}p_{\rm T}$ is the transverse-momentum distribution of associated ${\rm K}^0_{\rm S}$ or $\Xi^\pm$ hadrons within the selected correlation region. The relative production of strange and multi-strange hadrons is characterized through the $\Xi^\pm/{\rm K}^0_{\rm S}$ ratio. The associated ${\rm K}^{0}_{\rm S}$ and $\Xi^{\pm}$ hadrons are identified directly at generator level using their corresponding PDG particle IDs. No detector-level effect or weak-decay topology selection is applied. The selected particles are required to satisfy $|\eta|<4$, and their yields are integrated over the full generated transverse-momentum range down to $p_{\rm T}=0$. Unlike the experimental measurement, where the unmeasured low-$p_{\rm T}$ part of the spectra is extrapolated to obtain the $p_{\rm T}$-integrated yields, the generator-level calculation provides the complete particle spectrum and therefore requires no low-$p_{\rm T}$ extrapolation. Moreover, by construction, no acceptance correction through the event mixing technique is needed contrasting the generator-level analysis reported in the ALICE publication. The resulting yields thus correspond to the full generator-level $p_{\rm T}$-integrated production within the specified angular regions. Increasing $p_{\rm T}^{\rm trigg}$ progressively biases the event sample toward harder scattering processes, thereby providing a means to determine whether the toward-leading strangeness enhancement persists when contributions associated with softer particle production are reduced.\\

Events are classified into six multiplicity classes according to the percentile of the event-activity estimator constructed from the charged-particle activity in the forward and backward pseudorapidity regions (V0M-like multiplicity estimator as used by ALICE). Specifically, the activity is determined in the intervals \(2.8<\eta<5.1\) and \(-3.7<\eta<-1.7\), and the summed activity is used to define multiplicity percentiles following the procedure of Ref.~\cite{ALICE:2024zxp}. The six event classes correspond to the \(0\!-\!1\%\), \(1\!-\!5\%\), \(5\!-\!10\%\), \(10\!-\!30\%\), \(30\!-\!50\%\), and \(50\!-\!100\%\) percentiles of the event-activity distribution. For each event-activity class, the mean charged-particle density at midrapidity is evaluated after applying the trigger-particle requirement.  The charged-particle pseudorapidity density at midrapidity, \(\langle {\rm d}N_{\rm ch}/{\rm d}\eta\rangle\), is then evaluated within \(|\eta|<0.8\) for each multiplicity class. This separation between the event-activity estimator and the midrapidity region used for the physics observables minimizes autocorrelation effects and provides a consistent comparison with the ALICE analysis. The background-subtracted toward-region yield is obtained via a zero-yield-at-minimum (ZYAM) procedure; the near-side $\Delta\varphi$ profile of the full-region correlation, restricted to $|\Delta\eta|<0.86$ and rescaled to this window, is fitted with a constant in a narrow window $1.4<\Delta\varphi<1.6$ rad, taken to represent the flat underlying-event pedestal. This pedestal is then subtracted bin-by-bin from the raw toward-region $\Delta\varphi$ distribution and properly normalized with the $|\Delta\eta|$ range.

\subsection{Event Generators}

\subsubsection{PYTHIA8}
\begin{figure*}[ht]
    \centering
    \includegraphics[width=0.9\linewidth]{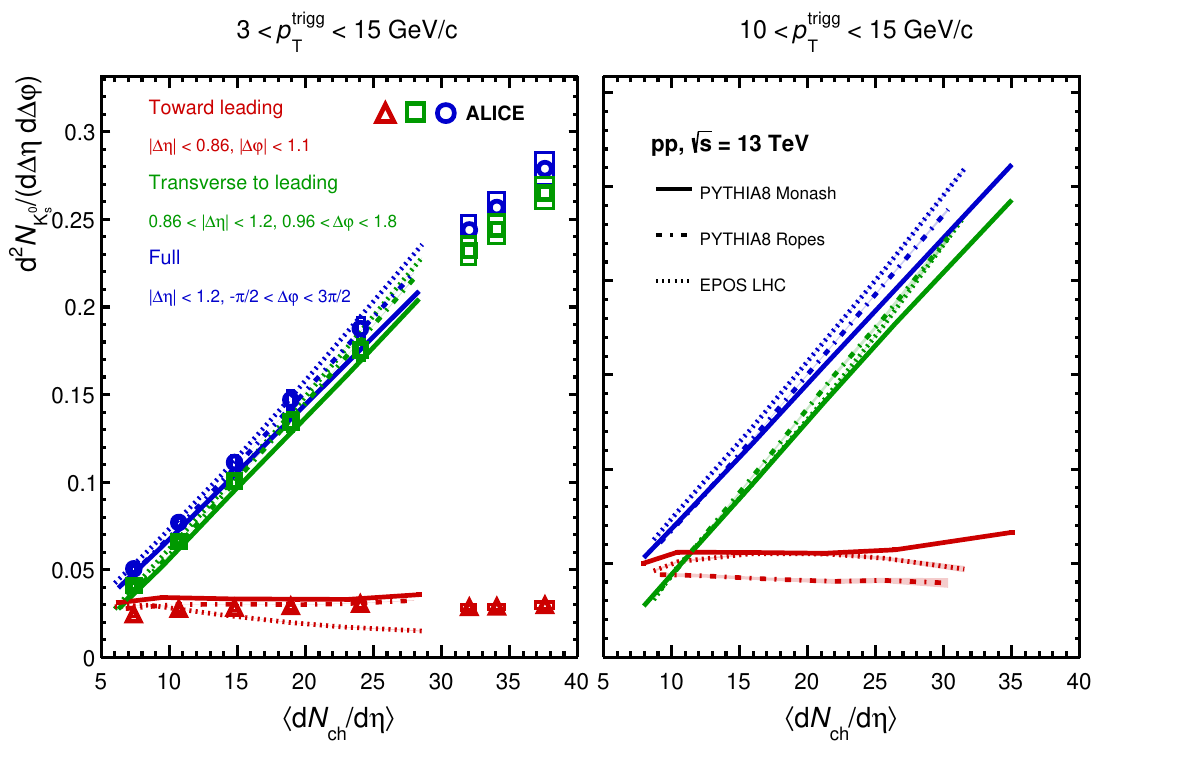}
    \caption{Per-trigger ${\rm K}^{0}_{\rm S}$ yields per unit $\Delta\eta\Delta\varphi$ area as a function of charged-particle multiplicity in the full (blue), transverse-to-leading (green), and toward-leading (red) regions. The left panel corresponds to the trigger selection $p_{\rm T}^{\rm trigg}>3$ GeV/$c$, while the right panel shows the results for $p_{\rm T}^{\rm trigg}>10$ GeV/$c$. The shaded bands represent the statistical uncertainty from the simulated event sample.}
    \label{fig1}
\end{figure*}

\begin{figure*}[t]
    \centering
    \includegraphics[width=0.9\linewidth]{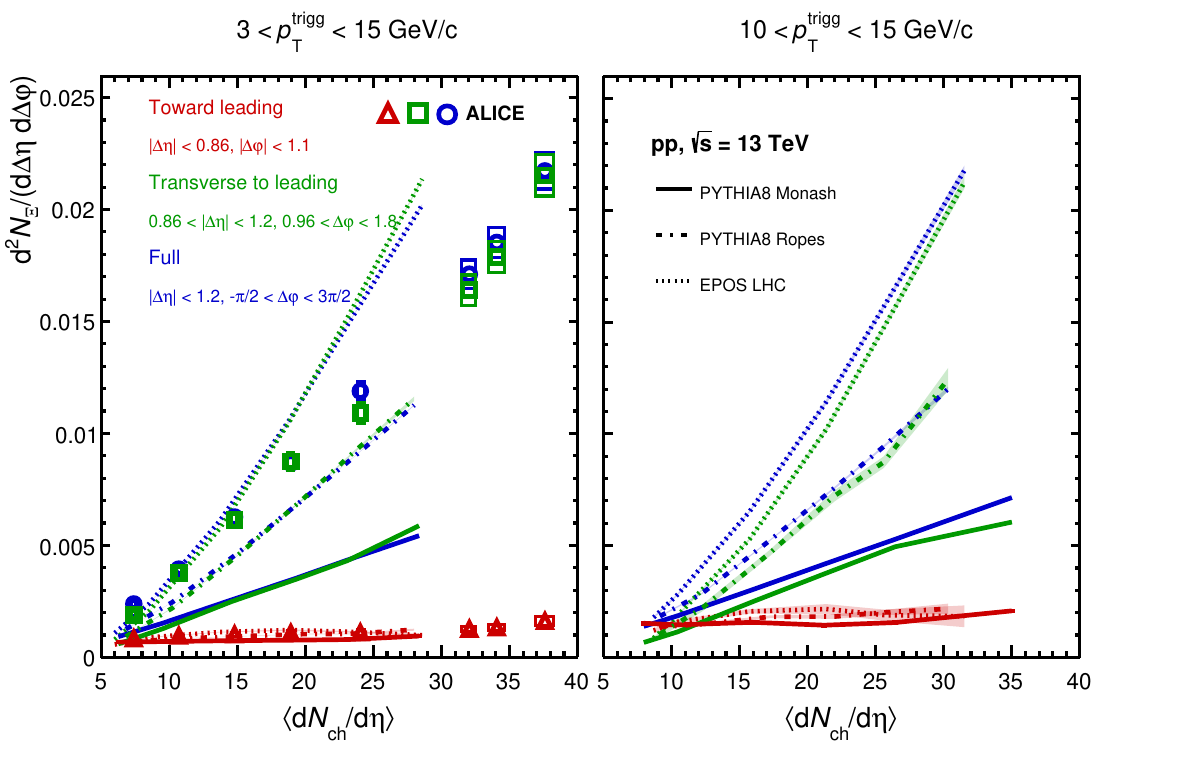}
    \caption{Per-trigger $\Xi^{\pm}$ yields per unit $\Delta\eta\Delta\varphi$ area as a function of charged-particle multiplicity in the full (blue), transverse-to-leading (green), and toward-leading (red) regions. The left panel corresponds to the trigger selection $p_{\rm T}^{\rm trigg}>3$ GeV/$c$, while the right panel shows the results for $p_{\rm T}^{\rm trigg}>10$ GeV/$c$. The shaded bands represent the statistical uncertainty from the simulated event sample.}
    \label{fig2}
\end{figure*}

To simulate ultrarelativistic hadronic collisions, event generators such as \textsc{PYTHIA8} provide a comprehensive framework for describing hard and soft QCD processes~\cite{Sjostrand:2014zea}. The generator incorporates hard scatterings through matrix elements, parton distribution functions, initial- and final-state parton showers, multiparton interactions (MPI), beam remnants, and hadronization. In particular, MPI provides an important contribution to the modeling of the underlying event and soft particle production accompanying hard scatterings~\cite{Corke:2009tk}.

Hadronization is described using the Lund string fragmentation model, which converts colored partons into color-neutral hadrons. Color reconnection allows alternative color connections between partons originating from different multiparton interactions and modifies the resulting string topology~\cite{Christiansen:2015yqa}. In addition to the standard hadronization framework, \textsc{PYTHIA8} can incorporate color-rope formation, in which overlapping strings combine coherently into higher-color representations with an enhanced effective string tension. This modifies the flavour composition of the produced hadrons and can enhance strange-hadron production~\cite{Bierlich:2014xba,Bierlich:2015rha}. In the present analysis, the Monash 2013 tune is employed with and without color-rope formation~\cite{Skands:2014pea}. The rope-free configuration provides a baseline for conventional Lund string fragmentation, while the rope configuration allows the role of local string-density effects in strangeness production to be investigated. The exact tunes for both PYTHIA8 Monash and ropes used in this study are given in the table~\ref{tab:pythia_settings}.

\subsubsection{EPOS-LHC}

EPOS-LHC is a Monte Carlo event generator based on the EPOS framework, which combines Gribov--Regge multiple-scattering theory with perturbative QCD to provide a unified description of soft and hard particle production in hadronic collisions~\cite{Pierog:2013ria}. The model treats a collision as a superposition of multiple parton ladders, with energy conservation enforced consistently at each interaction vertex. This approach provides a description of multiparton interactions and the associated underlying-event activity.

A distinctive feature of EPOS-LHC is the separation of the produced matter into a dense core and a dilute corona. Particles originating from the corona undergo conventional string fragmentation, whereas the dense core undergoes collective expansion before hadronization~\cite{Pierog:2013ria}. This treatment of collective dynamics allows EPOS-LHC to describe several features of particle production in high-multiplicity pp and p--Pb collisions, including enhanced strange-hadron production and flow-like signatures.

In the present work, EPOS-LHC is employed as a representative framework incorporating collective particle-production dynamics. Its simultaneous treatment of hard scatterings, multiparton interactions, and a collective core provides a complementary description to the local string-density mechanism implemented through color ropes in \textsc{PYTHIA8}. The comparison between the two frameworks therefore allows the dependence of strangeness enhancement on the trigger momentum to be investigated in the context of qualitatively different particle-production mechanisms.

\subsection{Event generation and selection}

All calculations are performed for pp collisions at
$\sqrt{s}=13~\mathrm{TeV}$. For PYTHIA8 Monash and ropes model, approximately 2.5 billion events are generated, whereas for EPOS-LHC case, 1 billion pp collisions were simulated. For \textsc{PYTHIA8}, the Monash tune is employed with and without color-rope formation. For EPOS-LHC, calculations are performed with parametrized hydrodynamic evolution. The same event and particle selections are applied to both frameworks, including the charged-particle multiplicity selection, the trigger-particle definition and the correlation regions.

\section{Results and Discussion}
\label{sec3}

\begin{figure*}[t]
    \centering
    \includegraphics[width=0.9\linewidth]{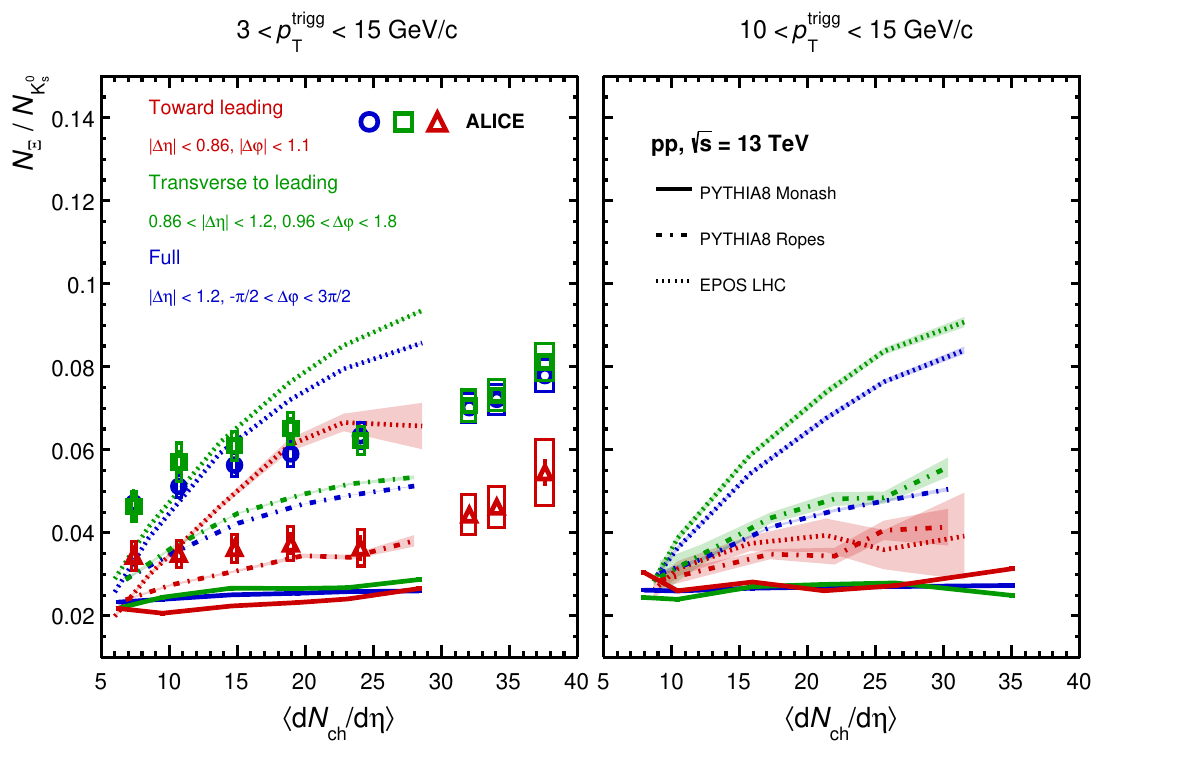}
    \caption{Multiplicity dependence of the per-trigger $\Xi^{\pm}/{\rm K}^{0}_{\rm S}$ yield ratio in the full (blue), transverse-to-leading (green), and toward-leading (red) regions. The left panel corresponds to the trigger selection $p_{\rm T}^{\rm trigg}>3$ GeV/$c$, while the right panel shows the results for $p_{\rm T}^{\rm trigg}>10$ GeV/$c$. The shaded bands represent the statistical uncertainty from the simulated event sample.}
    \label{fig3}
\end{figure*}

\begin{figure*}[t]
    \centering
    \includegraphics[width=0.9\linewidth]{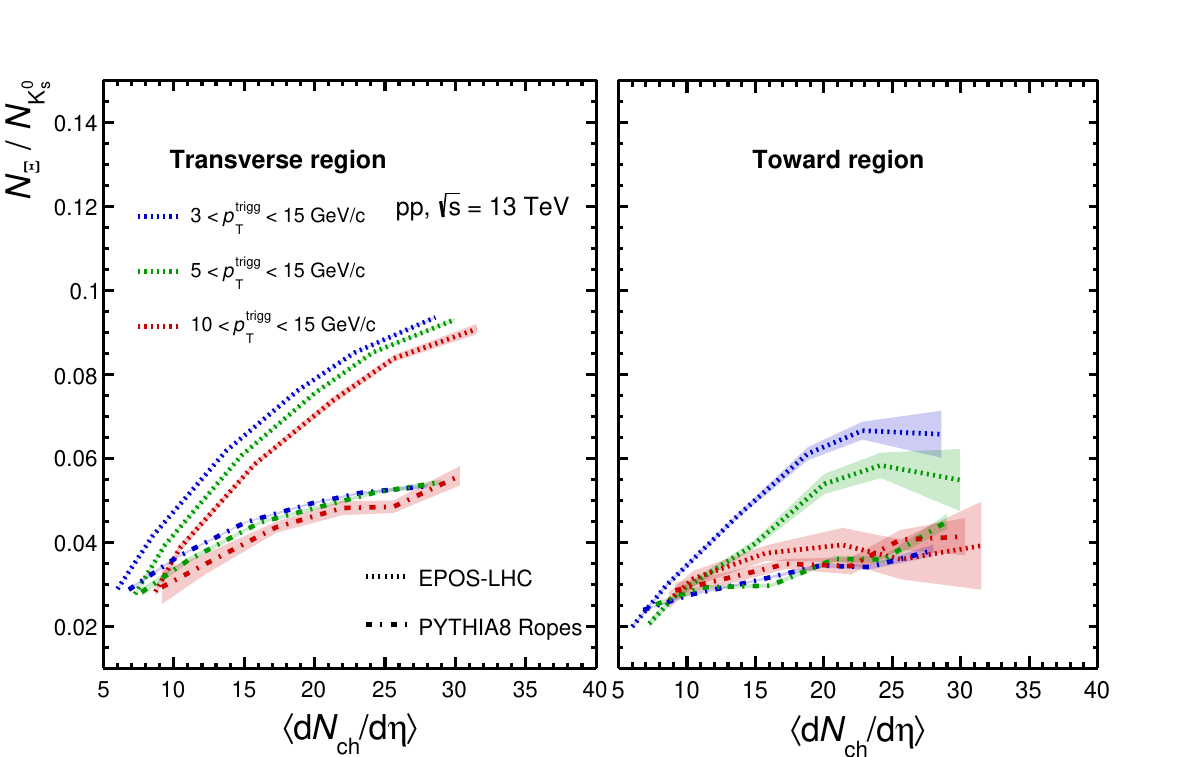}
    \caption{Multiplicity dependence of the per-trigger $\Xi^{\pm}/{\rm K}^{0}_{\rm S}$ yield ratio in transverse (left panel) and toward region (right panel) for EPOS-LHC and PYTHIA8 ropes for different momentum triggers. The shaded bands represent the statistical uncertainty from the simulated event sample.}
    \label{fig4}
\end{figure*}

\begin{figure*}[t]
    \centering
    \includegraphics[width=0.9\linewidth]{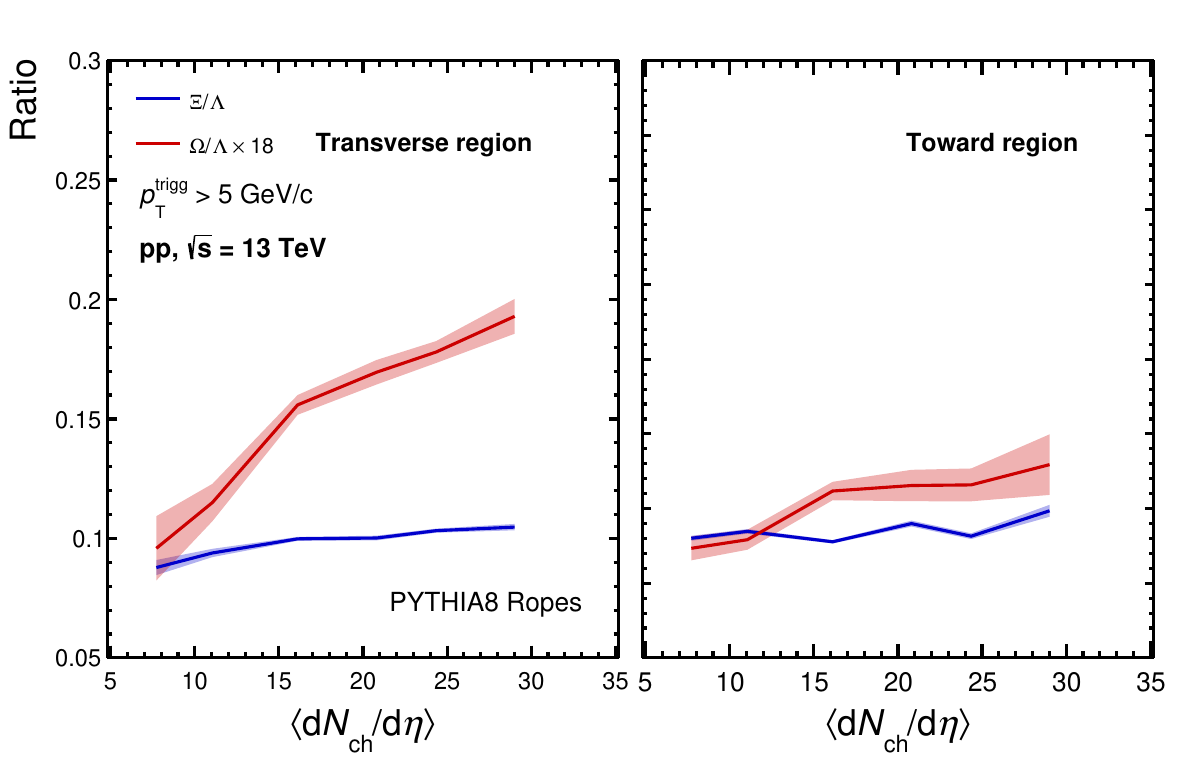}
    \caption{Multiplicity dependence of the per-trigger $\Xi/\Lambda$ and $\Omega/\Lambda$ yield ratio in transverse (left panel) and toward region (right panel) for PYTHIA8 ropes for $p_{\rm T}^{\rm trigg}>5$~GeV/$c$. The shaded bands represent the statistical uncertainty from the simulated event sample.}
    \label{fig5}
\end{figure*}

The left panel of Fig.~\ref{fig1} shows the multiplicity dependence of the per-trigger ${\rm K}^{0}_{\rm S}$ yield in the full, transverse-to-leading, and toward-leading regions for $p_{\rm T}^{\rm trigg} > 3$~GeV/$c$. The results from PYTHIA8 Monash, PYTHIA8 with ropes, and EPOS-LHC are compared with the ALICE data reported in Ref.~\cite{ALICE:2024zxp}. The yields shown here are full generator-level $p_{\rm T}$-integrated yields; the corresponding ALICE measurements are also reported as $p_{\rm T}$-integrated yields, with the experimentally inaccessible low-$p_{\rm T}$ region obtained through spectral extrapolation. A clear increase of the ${\rm K}^{0}_{\rm S}$ yield with charged-particle multiplicity is observed in both the full and transverse-to-leading regions, whereas the toward-leading yield is approximately flat. The transverse-to-leading region, which is less directly affected by the fragmentation of the leading jet, provides a sensitive probe of the underlying event~\cite{Ortiz:2018vgc}, while the toward-leading region receives a substantial contribution from jet fragmentation and therefore exhibits a stronger dependence on the trigger-$p_{\rm T}$ selection. The models reproduce the main qualitative trends reported in Ref.~\cite{ALICE:2024zxp}, although small quantitative discrepancies are observed due to the analysis-specific differences outlined above. The right panel of Fig.~\ref{fig1} presents the corresponding results for $p_{\rm T}^{\rm trigg} > 10$~GeV/$c$. Compared with the lower trigger-$p_{\rm T}$ selection, the difference between the full and transverse yields becomes more pronounced across all models. This indicates an enhanced contribution from the toward region for events with a harder trigger particle.

Similarly, the left panel of Fig.~\ref{fig2} shows the multiplicity dependence of the per-trigger $\Xi^{\pm}$ yield in the same three regions for $p_{\rm T}^{\rm trigg} > 3$~GeV/$c$, compared with the ALICE data. A clear increase of the $\Xi^{\pm}$ yield with charged-particle multiplicity is observed in the full and transverse-to-leading regions, indicating an enhanced production of multi-strange hadrons with increasing event activity, while the toward-leading yield shows a weaker multiplicity dependence. Differences among the particle-production scenarios become more pronounced for the $\Xi^{\pm}$ yield than for the ${\rm K}^{0}_{\rm S}$ yield, providing sensitivity to the treatment of strangeness production and hadronization. The right panel of Fig.~\ref{fig2} shows the corresponding results for $p_{\rm T}^{\rm trigg} > 10$~GeV/$c$.

The ratio of the strange baryon to strange meson, which can give information about the strangeness enhancement in the system, is now examined. The left panel of Fig.~\ref{fig3} shows the multiplicity dependence of the per-trigger $\Xi^{\pm}/{\rm K}^{0}_{\rm S}$ yield ratio in the full, transverse-to-leading, and toward-leading regions for $p_{\rm T}^{\rm trigg} > 3$~GeV/$c$, compared with the ALICE data. In the full and transverse-to-leading regions, the ratio exhibits a multiplicity dependence, indicating that the production of multi-strange hadrons increases relative to that of strange mesons with increasing event activity. In the toward-leading region, the multiplicity dependence is weaker, reflecting the larger contribution from jet fragmentation, but still a significant enhancement can be observed in both EPOS-LHC and PYTHIA8 (with ropes hadronization) as well as in the experimental data in particular at very high charged particle multiplicity. EPOS-LHC usually overestimates the data, but regardless, it shows a strong strangeness enhancement. On the other hand, PYTHIA8 Monash provides the medium-less baseline and stays flat as a function of charged particle multiplicity, while PYTHIA8 with ropes agrees with the data to a certain extent. The comparison between the different event generators highlights their sensitivity to the underlying particle-production and hadronization mechanisms. The right panel of Fig.~\ref{fig3} presents the corresponding results for $p_{\rm T}^{\rm trigg} > 10$~GeV/$c$. Here, it is observed that for the case of EPOS-LHC, the enhancement is strongly suppressed in the toward-leading case, whereas the transverse-to-leading case still shows the enhancement. However, a little to no change is seen for the PYTHIA8 with ropes result as compared to the low momentum trigger case.

Figure~\ref{fig4} shows the $\Xi^{\pm}/{\rm K}^{0}_{\rm S}$ yield ratio in transverse and the toward-leading region in the left and right panel respectively for three different $p_{\rm T}^{\rm trigg}$, where the results from EPOS-LHC to PYTHIA8 ropes are compared. To understand this behavior in detail, the trigger-momentum dependence of the \(\Xi^\pm/{\rm K}^{0}_{\rm S}\) ratio in both EPOS-LHC and PYTHIA8 with color ropes for three trigger-momentum intervals is studied: \(3<p_{\rm T}^{\rm trigg}<15\), \(5<p_{\rm T}^{\rm trigg}<15\), and \(10<p_{\rm T}^{\rm trigg}<15\) GeV/\(c\). While the transverse-to-leading region exhibits little dependence on $p_{\rm T}^{\rm trigg}$, a strong effect is observed in the toward-leading region. The EPOS-LHC prediction shows a systematic reduction of the multiplicity dependence of the ratio with increasing trigger momentum, whereas the PYTHIA8 rope prediction exhibits a substantially weaker dependence on the trigger selection. The different responses reflect the distinct particle-production mechanisms implemented in the two frameworks. In EPOS-LHC, the result is consistent with a reduced contribution from the dense collective core for increasingly hard trigger selections. In PYTHIA8 with color ropes, the weaker trigger dependence is consistent with the local string-overlap mechanism remaining active for harder trigger selections. The comparison therefore demonstrates that the trigger momentum provides sensitivity to the relative contributions of jet-associated and underlying-event particle production in the two model frameworks. However, the different predictions cannot be interpreted as a model-independent separation of collective and non-collective mechanisms, since the two generators differ simultaneously in their treatment of several aspects of particle production and hadronization. The trigger momentum can therefore serve as a useful discriminator between the different model responses and provides an experimentally testable handle on the mechanisms contributing to strangeness enhancement.

Finally, to deepen the understanding of the finite enhanced production of $\Xi$ with respect to ${\rm K}_{\rm S}^{0}$ as a function of charged particle multiplicity, the double-strange baryon $\Xi$ to single-strange baryon $\Lambda$ ratio is analyzed. In order to understand the systematic strangeness enhancement, this is compared with the ratio of $\Omega$, which is a triple-strange baryon, to $\Lambda$. In figure~\ref{fig5}, the baryon ratios are plotted for the case of $p_{\rm T}^{\rm trigg}>5$~GeV/$c$. The left panel shows the ratio in transverse region and the right panel shows the ratio in toward region. For a better visualization of the effect, the $\Omega$ to $\Lambda$ ratio is multiplied with a factor of 18 in both panels. It is observed that in the transverse region, the $\Omega/\Lambda$ yield ratio gives a systematic enhancement as a function of charged particle multiplicity as compared to $\Xi/\Lambda$ ratio. However, the weaker multiplicity dependence in the toward region indicates that the hierarchy of multi-strange-to-strange baryon production is less pronounced in the jet-like component than in the transverse region. This provides additional evidence that the multiplicity dependence of strangeness production reported in Ref.~\cite{ALICE:2024zxp} is sensitive to the relative contributions of jet-associated and underlying-event production.

\section{Summary}
\label{sec4}
The multiplicity dependence of strange-hadron production in high-multiplicity pp collisions is investigated using PYTHIA8 with and without color-rope formation, and EPOS-LHC. The study was performed for ${\rm K}^{0}_{\rm S}$ and $\Xi^{\pm}$ hadrons in the toward-leading and transverse-to-leading regions defined relative to the highest-$p_{\rm T}$ charged particle in the event, with particular emphasis on the role of the trigger-particle transverse momentum.

For the trigger selection employed in the recent ALICE measurement, $p_{\rm T}^{\rm trigg}>3$~GeV/$c$, both EPOS-LHC and PYTHIA8 with ropes predict a finite enhancement of the $\Xi^{\pm}/{\rm K}^{0}_{\rm S}$ ratio with increasing charged-particle multiplicity in both regions, whereas PYTHIA8 Monash is compared for baseline measurement. However, a qualitatively different behaviour emerges at $p_{\rm T}^{\rm trigg}>10$~GeV/$c$. While the enhancement remains significant in the transverse-to-leading region, it is strongly suppressed in the toward-leading region and becomes nearly independent of multiplicity. In EPOS-LHC, the toward-leading enhancement decreases for increasingly hard trigger selections, consistent with a reduced contribution from the collective core, whereas the PYTHIA8 color-rope prediction exhibits a substantially weaker trigger dependence, consistent with a local string-density mechanism that remains active for harder triggers. The persistence of the enhancement in the transverse-to-leading region, where the underlying-event contribution is larger, further emphasizes the important role of underlying-event particle production. These results show that the trigger momentum provides a sensitive probe of the interplay between jet-associated and underlying-event contributions to strange-hadron production, and offers a way to test the microscopic mechanisms responsible for strangeness enhancement in high-multiplicity pp collisions.\\

\section*{Acknowledgement}
Authors acknowledge the technical support from Luciano D\'iaz Gonz\'alez and Jes\'us Eduardo Murrieta Le\'on. This work has been supported by DGAPA-UNAM PAPIIT No. IG100524 and PAPIME No. PE100124.

\appendix
\section{PYTHIA8 generator settings}
\label{app:pythia_settings}

\begin{table*}[htbp]
\centering
\caption{PYTHIA8 generator settings used for the Ropes and Monash
configurations. Both configurations use inelastic $pp$ collisions at
$\sqrt{s}=13$~TeV with \texttt{SoftQCD:inelastic = on}.}
\label{tab:pythia_settings}
\renewcommand{\arraystretch}{1.15}
\begin{tabular}{llcc}
\toprule
\textbf{Category} & \textbf{Parameter} & \textbf{Ropes} & \textbf{Monash} \\
\midrule

Base tune
& \texttt{Tune:pp}
& -- (custom)
& 14 \\

\midrule
Fragmentation
& \texttt{StringPT:sigma}
& 0.335
& -- \\

\midrule
Multi-parton interactions
& \texttt{MultiPartonInteractions:pT0Ref}
& 2.15
& -- \\

\midrule
color reconnection
& \texttt{colorReconnection:mode}
& 1
& -- \\
& \texttt{colorReconnection:allowDoubleJunRem}
& on
& -- \\
& \texttt{colorReconnection:m0}
& 0.3
& -- \\
& \texttt{colorReconnection:allowJunctions}
& on
& -- \\
& \texttt{colorReconnection:junctionCorrection}
& 1.2
& -- \\
& \texttt{colorReconnection:timeDilationMode}
& 2
& -- \\
& \texttt{colorReconnection:timeDilationPar}
& 0.18
& -- \\

\midrule
Rope hadronization
& \texttt{Ropewalk:RopeHadronization}
& on
& off \\
& \texttt{Ropewalk:doShoving}
& on
& off \\
& \texttt{Ropewalk:doFlavour}
& on
& off \\
& \texttt{Ropewalk:tInit}
& 1.5
& -- \\
& \texttt{Ropewalk:deltat}
& 0.05
& -- \\
& \texttt{Ropewalk:tShove}
& 0.1
& -- \\
& \texttt{Ropewalk:gAmplitude}
& 0.0
& -- \\
& \texttt{Ropewalk:r0}
& 0.5
& -- \\
& \texttt{Ropewalk:m0}
& 0.2
& -- \\
& \texttt{Ropewalk:beta}
& 0.1
& -- \\

\midrule
Vertex information
& \texttt{PartonVertex:setVertex}
& on
& off \\

\midrule
Particle decays
& \texttt{ParticleDecays:limitTau0}
& on
& on \\
& \texttt{ParticleDecays:tau0Max}
& 10~mm
& 10~mm \\

\midrule

\bottomrule
\end{tabular}
\end{table*}

\end{document}